\documentclass[12pt,draft]{article}

\usepackage{amssymb}
\usepackage{latexsym}

\newtheorem{theorem}{Theorem}[section]
\newtheorem{definition}[theorem]{Definition}
\newtheorem{proposition}[theorem]{Proposition}
\newtheorem{lemma}[theorem]{Lemma}

\newcommand{\Z}{{\mathbb Z}}
\newcommand{\R}{{\mathbb R}}
\newcommand{\C}{{\mathbb C}}
\newcommand{\g}{{\mathfrak g}}
\newcommand{\dd}{{\rm d}}   
\newcommand{\tr}{{\rm tr}}  

\begin{document}

\title{Homotopic classification of Yang--Mills vacua\\
taking into account causality}

\author{
G\'abor Etesi
\\ {\it Department of Geometry, Mathematical Institute,}
\\ {\it Budapest University of Technology and Economics,}
\\{\it H \'ep., Egry J\'ozsef u. 1., Budapest,} 
\\{\it H-1111 Hungary}
\\ {\tt etesi@math.bme.hu}}

\maketitle

\pagestyle{myheadings}
\markright{G. Etesi: Causal classification of Yang--Mills vacua}

\thispagestyle{empty}

\begin{abstract}
Existence of $\theta$-vacuum states in Yang--Mills theories defined over 
asymptotically flat space-times examined taking into account not only the 
topology but the complicated causal structure of these space-times, too. 
By a result of Galloway apparently causality makes all vacuum states, seen
by a distant observer, homotopically equivalent making the
introduction of $\theta$-terms unnecessary. 

But a more careful analysis shows that certain twisted classical vacuum
states survive even in this case eventually leading to the
conclusion that the concept of ``$\theta$-vacua'' is meaningful in
the case of general Yang--Mills theories. We give a classification of
these vacuum states based on Isham's results showing that the Yang--Mills
vacuum has the same complexity as in the flat Minkowskian case hence the
general CP-problem is not more complicated than the well-known flat one.
We also construct the $\theta$ vacua rigorously via geometric 
quantization.
\end{abstract}
\vspace{0.1in}

\centerline{Keywords: {\it $\theta$-vacua; asymptotical flatness; 
causality; topological censorship}}
\centerline{PACS numbers: 11.15, 11.30.E, 04.20.G, 04.70}
\vspace{0.1in}

\section{Introduction: the Minkowskian Yang--Mills theory}
The famous solution of the long-standing $U(1)$-problem in the Standard 
Model via instanton effects was presented by 't Hooft about three
decades ago \cite{tho2}\cite{tho}. This solution demonstrated that {\it 
instantons} i.e., finite-action self-dual solutions of the {\it Euclidean}
Yang--Mills-equations
discovered by Belavin et al. \cite{bpst} should be taken seriously
in gauge theories. Another problem arose in these models over the
{\it Minkowskian} space-time, however: if instantons really exist, they 
induce a P- hence CP-violating so-called $\theta$-term in the effective 
Yang--Mills action. But according to accurate experimental results, such a
CP-violation cannot occur in QCD, for instance. The most accepted
solution to this  problem is the so-called {\it Peccei--Quinn mechanism}
\cite{pec}. A consequence of this mechanism is the existence of a light
particle, the so-called {\it axion}. This particle has not been observed
yet, however.

The question naturally arises whether or not such problematic
$\theta$-term must be introduced over more generic space-times. The aim of
our paper is to claim that the answer is yes.
 
First, let us summarize the vacuum structure of a gauge theory over 
Minkowski space-time following basic text books \cite{che}\cite{kak}. Let
$E$ be a complex vector bundle over an oriented and time
oriented Lorentzian manifold
$(M,g)$ belonging to a finite dimensional complex representation of $G$.
Without loss of generality we choose the gauge group $G$ to be a compact
Lie group. Consider a $G$-connection $\nabla$ on this bundle with
curvature $F_\nabla$; we take the usual Yang--Mills action (by fixing the
coupling to be $1$): 
\begin{equation}
S(\nabla ,g)=-{1\over 8\pi^2}\int\limits_M\tr\left(F_\nabla\wedge
*F_\nabla\right) ,
\label{eym}
\end{equation}
whose Euler--Lagrange equations are
\[\dd_\nabla F =0,\:\:\:\:\:\dd_\nabla *F=0.\]
Here $*$ is the Hodge operation induced by the orientation and the
metric on $M$. In our
present case $M=\R^4$ and usually the metric $g$ is fixed and supposed to
be the Minkowskian one on $\R^4$. Moreover all $G$-bundles $E$ are trivial
consequently by choosing a particular frame on $E$, the connection
$\nabla$ can be identified globally with a $\g$-valued $1$-form $A$. 

The simplest solution is the vacuum i.e., a flat connection: $F_\nabla
=0$. By simply connectedness of $\R^4$ such gauge fields can be written in
the form $A=f^{-1}\dd f$, where $f: \R^4\rightarrow G$ is a smooth
function.

But by the existence of a global temporal gauge on $\R^4$ (in this gauge 
flat connections are independent of the ``time'' variable) it is enough to 
consider the restriction of $f$ to a spacelike submanifold of Minkowski
space-time i.e., $f: \R^3\rightarrow G$. Minkowski space-time is
asymptotically flat as well, so there is a point $i^0$ called 
spacelike infinity. This point represents the ``infinity of space''
hence can be added to $\R^3$ completing it to the
three-sphere $\R^3\cup\{i^0\}=S^3$. It is well-known that vacuum fields
(possibly after a null-homotopic gauge-transformation around $i^0$) 
extend to the whole $S^3$ consequently classical vacua are classified by
maps $f: S^3\rightarrow G$. These maps up to homotopy are given by
elements of $\pi_3 (G)$. For typical compact Lie groups $\pi_3(G)$ is not 
trivial. This fact can be interpreted as classical vacua are separated 
from each other by energy barriers of finite height i.e., it is
impossible to deform two vacua of different winding numbers into each
other only through vacuum states. Hence homotopy equivalence reflects the
{\it dynamical structure} of the theory.

On the other hand, vacua are also acted upon by the gauge group. For 
simplicity assume $G\cong SU(2)$. In this case $\pi_3(SU(2))\cong\Z$. If 
$f_1$, $f_2$ are vacua of winding numbers $n_1$, $n_2$ respectively, there is a 
gauge transformation $g: S^3\rightarrow SU(2)$
of winding number $n_2-n_1$ satisfying $gf_1=f_2$. Consequently we can see
that the concept of {\it dynamical equivalence} of vacua reflecting the 
{\it dynamics} of the theory (i.e., the {\it homotopy equivalence} of maps 
$f: S^3\rightarrow SU(2)$) is different from that of {\it symmetry 
equivalence} of vacua representing the {\it symmetry} of the gauge theory 
(i.e., the {\it gauge equivalence} of the above maps). 

To avoid this discrepancy, we proceed as follows. Suppose we have
constructed the Hilbert space ${\cal
H}_{\R^4}$ of the corresponding quantum gauge theory. If $\vert
n\rangle\in{\cal H}_{\R^4}$ denotes the  quantum vacuum state belonging to
a classical vacuum $f$ of winding number $n$, the simplest way to
construct a state which is invariant (up to phase) under both dynamical
(i.e., homotopy) and symmetry (i.e., gauge)  equivalence is to 
formally introduce the quantum state
\begin{equation}
\vert\theta\rangle 
:=\sum\limits_{n=-\infty}^{\infty}{\rm e}^{{\bf i}n\theta}\vert 
n\rangle\:\in{\cal H}_{\R^4} ,\:\:\:\:\theta\in\R .
\label{teta}
\end{equation}
These formal sums are referred to as ``$\theta$-vacua''. 

From the physical point of view, the introduction of $\theta$-vacua is
also necessary. Although the vacuum states of different winding
numbers are separated classically, they can be joined semi-classically
i.e., by a tunneling induced by non-trivial instantons of the
corresponding {\it Euclidean} gauge theory.
Indeed, as it is well known, the $SU(2)$ instanton number is an element 
$k\in H^4(S^4, \Z )\simeq\Z$ (here $S^4$ 
is the one-point conformal compactification of the Euclidean flat $\R^4$. 
Note that the notion of ``instanton number'' comes from a very different
compactification compared with the derivation of ``vacuum winding
number''). If two vacua, $\vert n_1\rangle$, $\vert n_2\rangle$ ($n_1,
n_2\in \pi_3(SU(2))\simeq\Z$) are given then there is an instanton of instanton
number $n_2-n_1\in H^4(S^4,\Z )\simeq\Z$ tunneling between them in
temporal gauge \cite{che}\cite{kak}. In other words the true vacuum
states are linear combinations of the vacuum states of unique winding numbers 
yielding again (\ref{teta}). 

But the value of $\theta$ cannot be changed in any order of perturbation
i.e., it should be treated as a physical parameter of the theory; this
implies that tunnelings induce the effective term
\[{\theta\over 8\pi^2}\int\limits_{\R^4}\tr\left( F_\nabla\wedge
F_\nabla\right)\]
in addition to the action (\ref{eym}). But it is not difficult to see that
such a term violates the parity symmetry $P$ of the theory resulting in 
the violation of the CP-symmetry.

In summary, we have seen that there are at least three different ways to
introduce $\theta$-parameters in Yang--Mills theories {\it over
Minkowskian space-time}:

(i) $\theta$ is introduced to fill in the gap between the notions of
dynamical (i.e., homotopy) and symmetry (i.e., gauge)
equivalence of Yang--Mills vacua. This approach is pure mathematical in
its nature;

(ii) $\theta$ must be introduced because by instanton effects vacua of
definite winding numbers are superposed in the underlying semi-classical
Yang--Mills theory;

(iii) $\theta$ must be introduced by ``naturality arguments'' i.e.,
nothing prevents us to extend the Yang--Mills action at the full quantum 
level by a $P$-violating term $\tr\left( F_\nabla\wedge F_\nabla\right)$
with coupling constant $\theta$.

There is a correspondence between the above three characterizations of the
$\theta$ in {\it flat Min\-kow\-ski\-an space-time} but in the case of
general space-times, clear and careful distinction must be made until a
relation or correspondence between the three notions is established.
Clearly, (i) is related to the {\it topology} of the space-time and
the gauge group hence it is relatively easy to check whether or not it
remains valid in the general case. Concept (ii) is related to the
semi-classical structure of the
general Yang--Mills theory especially to the existence of instanton
solutions in the Wick-rotated theory and their relationship with vacuum
tunneling. The validity of concept (iii) is the most subtle one: we need
lot of information on the global non-perturbative aspects of a general
quantum Yang--Mills theory to check if any $\theta$-term survives quantum
corrections. In the present state of affairs, having no adequate general
theory of Wick rotation, instantons and their physical interpretation,
non-perturbative aspects of general Yang--Mills theories etc., we can
examine only the validity of concept (i) in the general case. Its validity
or invalidity may serve as a good indicator for the existence and role of
$\theta$-terms in general Yang--Mills theories.

The analysis of the vacuum structure of general Yang--Mills
theories over a space-time $(M,g)$ from the point of view of (i) was
carried out by Isham et al. 
\cite{isham5}\cite{isham1}\cite{isham3}\cite{isham4}. In
these papers Isham et al. argue that in the general case concept (i) for
introducing $\theta$-terms still continues to hold due to the complicated
topology of the spatial surface $S\subset M$ and the gauge group $G$
\cite{isham1}. The classical vacuum structure of these theories becomes
more complicated and we cannot avoid the introduction of various new
CP-violating terms into the effective Lagrangian \cite{isham5}. 

We have to emphasize that the approach of Isham et al. to the problem is
pure topological in its nature, however. By a result of Witt \cite{wit}
every oriented, connected three-manifold $S$ appears as a Cauchy surface
of a physically reasonable initial data set. It is well-known
that the complicated topology of the spacelike submanifold $S$ leads to
appearance of singularities in space-time if it arises as the
Cauchy development of $S$. Indeed, an early result of Gannon \cite{gan}
shows that the Cauchy development of a non-simply connected Cauchy surface
is geodesically incomplete i.e., singularities occur. If we accept the
Cosmic Censorship Hypothesis, these singularities are hidden behind event
horizons resulting in a non-trivial causal structure for these
space-times, too. A theorem of Galloway \cite{gal} (cf. an 
earlier version assumming stationarity by Chru\'sciel--Wald
\cite{chr-wal}) shows that distant observers can observe only
simply connected portions of 
asymptotically flat space-times: all topological properties are hidden
behind event horizons, eventually resulting again in a topologically
simple {\it effective} space-time. Hence one may doubt if Isham's
conclusions remain valid.

In Section 2 we formulate Yang--Mills theories with 
an arbitrary compact gauge group over general asymptotically
flat space-times satisfying the null energy condition with a 
single globally hyperbolic domain of outer communication. This model 
provides a good framework for analysing classical Yang--Mills vacua  over 
causally non-trivial space-times. In this setup we simply mimic the above 
analysis concerning classical Yang--Mills vacua and find that although
all vacua are topologically equivalent on the causally connected 
regime of the space-time, the appearance of a natural boundary condition
on the event horizon (also a consequence of the causal structure) 
introduces non-trivial homotopy classes again. 

In Section 3 we calculate explicitly the homotopy classes of vacua for
the classical groups. A modification appears compared with Isham and
other's pure topological considerations in the sense that
generally the vacuum structure in our case has exactly the same 
complexity as in the flat Minkowskian case, a surprising result. This 
demonstrates the ``stability'' of the $\theta$-problem and justifies 
concept (i) even in the more general case.

The idea of studying relationship between micro- or virtual black
holes, wormholes and $\theta$-vacua is not new. For example, see 
Hawking \cite{haw} and Preskil et al. \cite{pres-triv}. An earlier,
still incomplete version of this paper appeared in \cite{ete}.

\section{Asymptotically flat Yang--Mills theory}
The general reference for this chapter is \cite{wal}. Let $(M,g)$ be 
a four dimensional, oriented and time oriented smooth Lorentzian manifold 
i.e., a space-time; choose a complex vector bundle $E$ over $M$ associated 
to a principal bundle with compact gauge group $G$ via a finite 
dimensional complex representation. Consider a $G$-connection $\nabla$ on 
$E$ and a Yang--Mills theory with action (\ref{eym}) over $(M,g)$. We will 
focus on {\it vacuum solutions on a gravitational background} i.e.,
pairs $(\nabla , g)$ where $\nabla$ is a smooth flat $G$-connection on the
bundle $E$ while $g$ is a smooth Lorentzian metric on $M$. We will
suppose that $g$ is a solution of the vacuum or the coupled Einstein's 
equation with a matter field given by a stress-energy tensor $T$ obeying 
the null energy condition. We will refer the 
collection $(E, \nabla , M, g)$ to as an {\it Yang--Mills vacuum setup}.

We impose two restrictions. First, we will assume that $(M, g)$
contains a single {\it asymptotically flat region}. At a first look (for the
precise definitions see e.g. \cite{wal}) this means
that there is a conformal embedding $i: (M, g)\rightarrow
(\widetilde{M},\widetilde{g})$ such that the infinitely distant points
of $M$ are represented by the connected set $\partial\overline{i(M)}$ 
in the inclusion; furthermore this set is divided naturally into
three subsets: the future and past null infinities ${\cal I}^\pm$ and
the spatial infinity $i^0$. We remark that $\widetilde{g}$ is not supposed
to be smooth in $i^0$, even if $(M,g)$ is smooth.

Now consider the {\it domain of outer communication}
$\widetilde{N}\subseteq\widetilde{M}$ 
defined as $\widetilde{N}:=J^-({\cal I}^+)\cap J^+({\cal I}^-)$ and
$N:=\widetilde{N}\cap i(M)$. (Here $J^\pm
(X)$ denote the causal future and past of a subset $X$ in a space-time,
respectively). Notice that $N=M\setminus (B\cup W)$ where $B$ and $W$
are the black hole and white hole regions of $M$, respectively. The
boundary $\partial (B\cup W)$ is called the {\it event horizon} of these
regions. Our second assumption is that $(N, g\vert_N)$ is {\it
globally hyperbolic}. Consequently $N\cong S\times\R$ with $S$ being a
Cauchy surface for the domain of outer communication $N$ such that the
image of the Cauchy surface can be completed to a maximal spacelike
submanifold $\widetilde{S}$ in $\widetilde{M}$ by adding the spacelike
infinity $i^0\in\widetilde{M}$ to it: $i(S)\cup\{ i^0\} =\widetilde{S}$. 

Before proceeding further we fix notation. Let $V$ be a smooth, compact, 
oriented three-manifold (possibly with non-empty boundary),
$x^0\in V\setminus\partial V$ and assume there is a
homeomorphism $\varphi : V\setminus\partial V\rightarrow\widetilde{S}$
such that $\varphi (x^0)=i^0$. In this case we will say that {\it $S$ is
homeomorphic to the interior of $V$}. By global hyperbolicity, there is a
global time function $T: N\rightarrow\R$. Let $S_t:=T^{-1}(t)$ ($t\in\R$)
be a Cauchy surface which is the interior of a compact three-manifold $V$
(notice that $S_t\cong S_{t'}$ for all
$t,t'\in\R$). Consider a map $\varphi_t:V\setminus
(\partial V \cup\{x^0\})\rightarrow N$ whose image is
$\varphi_t(V\setminus (\partial V\cup\{x^0\}))=S_t\subset N$. The 
points $\varphi_t(x)=(x,t)$ of $S_t$ will be denoted as $x_t$. Clearly $V$ 
represents the compactification of a particular Cauchy surface 
since $V\cong\overline{\varphi_t^{-1}(S_t)}\cong
\overline{i(S_t)\cup\{i^0\}}$ for all $t\in\R$. Therefore by abuse of 
notation we will often think $S_t\subset V$ for all $t\in\R$.

Now we are in a position to address the problem of describing the topology 
of Yang--Mills vacua {\it seen by an observer in the domain of outer 
communication} of the space-time $(M,g)$. Clearly, at least classically, 
only this part of the space-time can be relevant for ordinary macroscopic 
observers. To achieve our goal, we refer to a general result of Galloway 
\cite{gal} (for an earlier version assuming stationarity cf. Chru\'sciel 
and Wald \cite{chr-wal}).
\begin{theorem} {\em (Galloway, 1995)}. {\it Let $(M, g)$ be an
asymptotically
flat space-time containing a single asymptotically flat region whose
domain of outer communication $(N, g\vert_N)$ is globally hyperbolic.
Suppose that the null energy condition holds. 

Then $N$ is simply connected i.e., $\pi_1(N)=1$.

Assume there is a Cauchy surface $S_t$ of $N$
homeomorphic to the interior of a compact three-manifold
$V$. Then if $\partial V\not=\emptyset$, each connected
component of $\partial V$ is homeomorphic to $S^2$.} $\Diamond$
\label{tetel}
\end{theorem}
This rather surprising observation is a consequence of the so-called 
Topological Censorship Theorem of Friedman--Schleich--Witt 
\cite{fri-sch}. 

We can see that $V$, the 
compactification of a Cauchy surface $S_t$ for $N$, is a simply 
connected (hence orientable) three-manifold. If $M$ contains black or 
white hole domains then $\partial V\not=\emptyset$  and all boundary 
components are homeomorphic to a two-sphere $S^2$ (``the event horizon of 
a black or white hole in an asymptotically flat space-time has no 
handles'').

The following simple lemma ensures us that from a technical 
viewpoint the vacuum structure at least over the relevant part $(N, 
g\vert_N)$ is exactly the same as in the Minkowskian case.
\begin{lemma}
Let $(M,g)$ be a space-time as in Theorem \ref{tetel} and $(E,\nabla
,M,g)$ be a Yang--Mills vacuum setup over it. Consider
the domain of outer communication with the
restricted Yang--Mills data $(E\vert_N, \nabla\vert_N, N, g\vert_N)$. 
Then

(i) If $\nabla\vert_N$ is flat and smooth then it can be identified with a
$\g$-vauled 1-form $A$ over $N$ and there is a smooth function
$f:N\rightarrow G$ such that $A=f^{-1}\dd f$;

(ii) There is a smooth gauge transformation $g: N\rightarrow G$
transforming $\nabla\vert_N$ into temporal gauge i.e., there is an
$A'=gAg^{-1}+g\dd g^{-1}$ such that $A'_0=0$ where $A'_0=A'(${\rm
grad}$T)$. If $A'$ is flat then the corresponding $f$ does not depend on 
$t$;

(iii) Fix a $t\in\R$ and consider the restriction $f\vert_{S_t}=:f_t: 
S_t\rightarrow G$. Then $f_t$ extends smoothly across the spacelike 
infinity $i^0$ i.e., there is a smooth function  
$\tilde{f}_t:\widetilde{S}_t\rightarrow G$, homotopic to $f_t$ on $S_t$. 
\label{allitas}
\end{lemma}
{\it Proof.} Concerning (i), the restricted bundle $E\vert_N$ is trivial 
hence any $G$-connection on it can be identified with a ${\mathfrak 
g}$-valued 1-form $A$; simply connectedness of $N$ implies that any
flat connection $\nabla\vert_N$ must be the trivial connection hence in
any gauge it can be represented in the form $A=f^{-1}\dd f$ as claimed.

To see (ii) we can write down the required gauge transformation by
solving the ordinary differential equation
\[gA_0g^{-1}+g{\partial g^{-1}\over \partial t}=0\]
over $N\cong S\times\R$. The solution over a chart $U\subset S$ is
\[ g(x,t)={\rm exp}\left(\int\limits_0^tA_0(x,\tau )\dd\tau\right)\]
with $x\in U$, $t\in\R$ and exp: $\g\rightarrow G$ being the
exponential map. This solution exists for finite $t$'s. 

The case of part (iii) is also very simple. Notice that there is a
neighbourhood $U\subset\widetilde{S}_t$ of $i^0$ such that $U\setminus\{
i^0\}\cong S^2\times [0,1)$. Consider the restriction
$f_t\vert_{S^2\times\{ 0\}}$ and take the function id: $S^2\times
[1/2 , 0)\rightarrow G$ sending all elements to the unit $e\in G$.
Then, taking into account that $\pi_2 (G)=0$ for compact Lie groups, there
is a smooth homotopy from $S^2$ to $G$ along $S^2\times [0, 1/2]$
connecting $f_t\vert_{S^2\times\{ 0\}}$ with id$\vert_{S^2\times\{
1/2\}}$. But this deformed function $\tilde{f}_t$ extends as the identity
to the whole $\widetilde{S}_t$ and is homotopic to $f_t$ on $S_t$.
$\Diamond$
\vspace{0.1in}

\noindent A pure Yang--Mills theory being conformally invariant, we may 
consider our Ein\-stein-matter theory together with a Yang--Mills field 
over $(\widetilde{M}, \widetilde{g})$ instead of the original space-time. 
The restriction of the extended flat Yang--Mills bundle $\widetilde{E}
\vert_{\widetilde{N}}$ is trivial even in this case. Certain physical
quantities of the extended theory may suffer from singularities on the
boundary $\partial\overline{i(M)}$ but classical Yang--Mills vacua in
temporal gauge 
extend smoothly over the whole $(\widetilde{M}, \widetilde{g})$ as we
have seen by the above lemma. In other words the studying of the
vacuum sector of the extended Yang--Mills theory is correct.

Summing up, we can see that dynamically (i.e., homotopically) inequivalent
vacua of the Yang--Mills theory are classified by the homotopy classes of
smooth maps $f: V\rightarrow G$ satisfying $f(i^0)=e\in G$, usually
written as  
\begin{equation}
\left[( V, i^0), (G, e)\right].
\label{vakuum}
\end{equation}
Now suppose that $(M, g)$ contains black and white hole(s). In this case
$V$ is a simply connected compact three manifold {\it with boundary} by
the theorem of Galloway. 
Such manifolds, considered as CW-complexes, have only cells of dimension
less than three. Hence by the Cellular Approximation Theorem \cite{spa},
every map $f: V\rightarrow G$ descends to a homotopic
map with values only on the cells of $G$ having dimension less than three.
Being $\pi_2(G)=0$, $G$ can be approximated by the simple
Postnikov-tower $P_2=K(\pi_1(G), 1)$ where $K(\pi_1(G), 1)$ is an
Eilenberg--Mac Lane space yielding
\begin{equation}
\left[( V, i^0), (G,e)\right]\cong
\left[ V,K(\pi_1(G), 1)\right]\cong H^1(V, \pi_1(G))=0.
\label{szamolas}
\end{equation}
The result is zero because $V$ is simply connected. For
details, see for instance \cite{spa}. Consequently all vacuum states are
homotopy-equivalent i.e., can be deformed into each other only through
vacuum states {\it over the domain of outer communication $N$ of
the space-time $(M,g)$}. Clearly, classically only this part is relevant
for a distant observer. 

This result can be explained from a different point of view as well. Since
the outer part $N$ of $M$ is globally hyperbolic by assumption, the
spacelike submanifold $S$ forms a Cauchy surface for $N$. Consequently if
we know the initial values of two gauge fields, $A$ and  $A'$ say, on
$S\subset N$, we can determine their values over the whole {\it outer}
space-time $N\subset M$ by using the field equations. This implies that
the values of the fields $A$ and $A'$ ``beyond'' the event horizon in a
moment are irrelevant for an observer outside the black hole. But we just
proved that every vacuum fields restricted to $V\supset S$ are homotopic.
Roughly speaking, homotopical differences between Yang--Mills vacua ``can
be swept'' into a black hole. 

Via (\ref{szamolas}) for arbitrary smooth functions
$f,g: V\rightarrow G$ there is a homotopy
\begin{equation}
F_T: V\times [0,1]\rightarrow G
\label{T-homotopia}
\end{equation}
satisfying $F_T(x,0)=f(x)$ and
$F_T(x,1)=g (x)$ and $F_T(i^0, t)=e$ for all
$(x,t)\in V\times [0,1]$.
Taking two Cauchy surfaces $T^{-1}(t_0)=:S_0$ and
$T^{-1}(t_1)=:S_1$ we can regard the two functions as vacua
$f\vert_{S_0}:=f_0 :S_0\rightarrow G$ and $g\vert_{S_1}:=f_1: 
S_1\rightarrow G$. In the 
homotopy $F_T$ the subscript ``$T$'' shows that the ``time'' required for the
homotopy is measured by the time function $T$ naturally associated to the
globally hyperbolic space-time $(M,g)$.

But on physical grounds, such a deformation or homotopy is effective 
only if the vacuum states, corresponding to the inital and final stages
of the homotopy, can be compared by an observer in finite proper time.
This means the following. Let $k=0,1$ and for all $x_k\in S_k$ for which
$f_0(x_0)\not=f_1(x_1)$ there must
exist an observer $\gamma :\R\rightarrow N$ moving forward in the region
$N$ who can measure hence compare $f_0(x_0)$ and $f_1(x_1)$ i.e., there
are $\tau_k\in \R$ such that a future directed light beam starting from
$x_k$ meets $\gamma$ in $\gamma (\tau_k)$, and
the proper time between $\gamma (\tau_0)$ and $\gamma (\tau_1)$ measured
by $\gamma$ is finite. In other words, there is a $\tau^-\in\R$ such
that $C\subset J^-(\gamma (\tau^-))$ where $C\subset S_0\times S_1$ 
contains the set of all points where $f_0(x_0)\not=f_1(x_1)$ with $x_k\in 
S_k$. Because our space-time may contain white hole regions too, we require the 
existence of another $\tau^+<\tau^-$ satisfying 
$C\subset J^+(\gamma (\tau^+))$ as well. The formal definition of 
such ``effective'' or ``observable'' homotopies is the following.
\begin{definition}
Let $(M,g)$ be an asymptotically flat space-time with a single globally
hyperbolic domain of outer communication $(N, g\vert_N)$ and let $T:
N\rightarrow\R$ be an associated time-function. Consider a homotopy of the
form (\ref{T-homotopia}) and let $C\subset S_0\times S_1$ be such that
$f_0(x_0)\not=f_1(x_1)$ with $(x_0,x_1)\in C$. 

Then (\ref{T-homotopia}) 
is called an {\em effective homotopy} or {\em observable homotopy} if 
there is a future directed non-spacelike piecewise 
smooth curve $\gamma :\R\rightarrow N$ and fixed numbers $\tau^\pm\in\R$ 
such that $C\subset J^\pm (\gamma (\tau^\pm))$.
\label{definicio}
\end{definition} 
{\it Remark.} We can see that in Minkowski space-time all
homotopies of the form (\ref{T-homotopia}) are effective homotopies
establishing the structure of the Minkowskian Yang--Mills vacuum also from
a physical viewpoint.

The following lemma is straightforward.
\begin{lemma}
Let $(M,g)$ be a space-time as in Definition \ref{definicio}
and consider a continuous curve $x: [0,\varepsilon ]\rightarrow V$ 
satisfying $x(0)\in \partial V$. Then we have induced spacelike curves 
$x_k: [0,\varepsilon ]\rightarrow S_k$ ($k=0,1$) given by 
$x_k(s)=(x(s),k)$ and satisfy $x_k(0)\in H$. 

The (abstract) homotopy (\ref{T-homotopia}) is effective if and only 
if there is a $0<\delta$ such that 
\[F_T(x_0(s), t)=F_T(x_1(s), t)\]
for all $0\leq s<\delta<\varepsilon$ and $t\in [0,1]$ that is, the 
homotopy is trivial in the vicinity of $H$. 
\label{propozicio}
\end{lemma}
{\it Proof.} We rerstrict our attention first to the case
$H^+=\partial B$, the future 
event horizon of the black hole regime $B$. Take a
homotopy of the form (\ref{T-homotopia})
with $F_T(x_k(s),k)=f_k(x_k(s))$ ($k=0,1$) and assume $F_T$ is
effective. By construction $x_k(0)\in H^+$ and if 
$f_0(x_0(0))\not=f_1(x_1(0))$ then there must exist a future
directed non-spacelike curve $\gamma :\R\rightarrow N$ such that
$\{x_0(0),x_1(0)\}\subset J^-(\gamma (\tau^-))$ for some
$\tau^-\in\R$. However this contradicts the definition of the domain of
outer communication $N$ consequently we must have $x_k(0)\notin H^+$. 
We get the same result for the past white hole horizon $H^-=\partial W$.
Therefore $x_k(0)\notin H=H^+\cup H^-$ as claimed. $\Diamond$
\vspace{0.1in}

\noindent From here we can see that given an abstract
homotopy (\ref{T-homotopia}), it gives rise to an effective homotopy if
and only if $F_T$ is {\it constant along $H$}. This result can be 
interpreted as a natural boundary condition on each
connected component of $\partial V$ for effectively deformable vacua 
dictated by the causal structure.
Since each boundary component in a ``moment'' is homeomorphic to the
two-sphere $S^2$ and $\pi_2(G)=0$ we can extend $f_0, f_1$ within their 
homotopy classes in the spirit of part (iii) of Lemma \ref{allitas} to 
functions $f,g:V\rightarrow G$ obeying $f(\partial V)=g(\partial V)=e\in 
G.$ The same argumentation yields the conditions 
$f(i^0)=g(i^0)=e$. We just remark that exactly this is the physical reason 
for keeping the functions as identity in spacelike infinity $i^0$ when we 
discuss homotopy classes of vacua over Minkowskian space-time: the 
spacelike infinity is invisible for an observer in $N$.

Therefore {\it the classes 
of effectively deformable vacua} are given by the homotopy classes of
functions $f: V\rightarrow G$ with the property
$f(\partial V)=f(i^0)=e\in G$. The homotopy is also
restricted to obey these boundary conditions. This set is denoted by
\begin{equation}
\left[ (V, \partial V, i^0), (G, e)\right]
\label{ujvakuum}
\end{equation}
and replaces (\ref{vakuum}). To get a more explicit description of
this set, we proceed as follows.

\section{Homotopic classification}
First taking into account that a function
$f: V\rightarrow G$ we are interested in satisfies
that it sends each connected component of $\partial V$ into the
unit element $e\in G$, we can replace the simply connected, compact
three-manifold-with-boundary $V$ with a closed,
simply connected three-manifold $W$ in the following way.
Let us denote by $k>0$ the number of connected components of
$\partial V$ (i.e., the number of black holes and white holes). As we have
seen, all such component is an $S^2$. Hence we can glue to each such
component a three-ball $B^3$ using the identity function of $S^2$ to get a
three-manifold without boundary
\[W:=V\cup_{\partial V}\underbrace{B^3
\cup\dots\cup B^3}_k.\]
Clearly, $f$ extends as the identity to each ball giving rise to
the function $f: W\rightarrow G$. Consequently, if we fix a point $x^n$
in each ball ($n\leq k$), then we may equivalently consider functions
obeying $f(x^1)=\dots =f(x^k)=f(i^0)=e$. Modifying the
allowed homotopies to obey this constraint, we can replace the homotopy
set (\ref{ujvakuum}) by
\[\left[ (W, x^1,\dots ,x^k, i^0), (G,e)\right]\]
(of course if $k=0$ then no point except $i^0$ is distinguished in
$W$). We prove the following proposition:
\begin{proposition}
Fix a number $k>0$ and consider the
connected, closed, simply connected three-manifold with $k+1$ distinguished 
points $(W, x^1,\dots ,x^k, i^0)$ constructed above. Denote by
$(W, i^0)$ the same space with only one distinguished point.
Then there is a natural bijection
\[\left[ (W, x^1,\dots ,x^k, i^0), (G,e)\right]\cong\left[ (W, i^0),
(G,e)\right]\]
by forgetting the points $x^1,\dots ,x^k\in W$ and modifying the allowed 
homotopies accordingly.
\label{homotopiapropozicio}
\end{proposition}
{\it Proof.} Fix a number $k\geq 0$. First it is straightforward that if
two functions, $f_0$ and $f_1$ are homotopic in $\left[
(W, x^1,\dots ,x^k, i^0), (G,e)\right]$ then they represent
the same homotopy class in $\left[ (W, i^0), (G,e)\right]$
i.e., they are homotopic in the later set as well. This is because the
allowed homotopies in $\left[ (W, i^0), (G,e)\right]$ are
less restrictive than in $\left[ (W, x^1,\dots ,x^k, i^0),
(G,e)\right]$. 

Conversely, it is not difficult to see that in each class
$[f]\in\left[ (W, i^0), (G,e)\right]$
there is a representant which belongs to $\left[ (W, 
x^1,\dots ,x^k, i^0), (G,e)\right]$. Indeed, choose an arbitrary
representant
$f\in [f]\in\left[ (W, i^0), (G,e)\right]$
and consider the pre-image $f^{-1}(e)\subset W$. This
pre-image contains the point $i^0\in W$ by construction.
Taking into account that $W$ is path connected, we can 
deform $f^{-1}(e)$ to contain the points $x^1,\dots ,x^k$ as well
producing a representant which belongs to $\left[ (W, x^1,\dots ,x^k,
i^0), (G,e)\right]$. 

Now suppose that there are two functions $f_0$ and $f_1$
which are homotopic in $\left[ (W, i^0), (G,e)\right]$ i.e.,
there is a continuous function $F: (W, i^0)\times
[0,1]\rightarrow (G,e)$ with
\[F(x, 0)=f_0(x),\:\:\:F(x,1)=f_1(x),\:\:\:F(i^0,
t)=e\:\:\:\:\:\mbox{for all $t\in [0,1]$ and $x\in (W, i^0)$}.\]
For the sake of simplicity, assume they represent elements in $\left[
(W, x^1,\dots ,x^k, i^0), (G,e)\right]$, too. Then we have to
prove that they are also homotopic in $\left[ (W, 
x^1,\dots ,x^k, i^0), (G,e)\right]$ i.e., there is a function $F':
(W, x^1,\dots ,x^k, i^0)\times 
[0,1]\rightarrow (G,e)$ with the property
\[F'(x,0)=f_0(x),\:\:\:F'(x,1)=f_1(x),\:\:\:F'(x^1,t)=\dots 
=F'(x^k,t)=F'(i^0,  t)=e\]
for all $t\in [0,1]$ and $x\in (W, x^1,\dots ,x^k, i^0)$.
From here we can see that the orbit of an arbitrary distinguished point
$x^n$ is a loop $l^n: [0,1]\rightarrow G$ under the homotopy $F$ while
the constant loop in the case of $F'$. Hence if these loops are
homotopically trivial in $G$ then we can deform $F$ into the homotopy $F'$
by shrinking the loops $l^1$,\dots ,$l^k$.

Now we prove that this is always possible. First, if $\pi_1(G)=1$ i.e., 
the compact Lie group is simply connected then certainly each loop $l^n$
is homotopic to the constant loop. Consequently assume $\pi_1(G)\not=1$.
Consider a distinguished point $x^n\in W$ and two paths $a^n:
[0, 1/2]\rightarrow W$ with $a^n(0)=i^0$ and $a^n(1/2)=x^n$
and $b^n: [1/2, 1]\rightarrow W$ with $b^n(1/2)= x^n$ and
$b^n(1)=i^0$. These give rise to a continuous loop $b^n*a^n
:[0,1]\rightarrow W$ with $b^n*a^n(0)=b^n*a^n(1)=i^0$. Here $*$ refers
to the juxtaposition of curves, loops, etc. 
The loop $b^n*a^n$ is homotopic to the trivial loop since $W$ is
simply connected. Consider the maps $\alpha_0^n:=f_0\circ a^n: [0,
1/2]\rightarrow G$ and $\beta_0^n:=f_0\circ b^n: [1/2,
1]\rightarrow G$. These are loops in $G$ hence so is their product
$\beta_0^n*\alpha_0^n$. Construct the same kind of loops
$\alpha_1^n:=f_1\circ a^n$ and $\beta_1^n:=f_1\circ b^n$.
The product loop $\beta^n_1*\alpha^n_1$ is homotopic in $G$ to
$\beta_0^n*\alpha_0^n$ i.e., 
$[\beta_0^n*\alpha_0^n]=[\beta^n_1*\alpha^n_1]$ because $f_0$ is
homotopic to $f_1$. It is clear that 
\[\beta^n_1*\alpha^n_1 =\beta_0^n*l^n*\alpha_0^n.\]
Consequently we can write for the homotopy classes in question
\[[\beta^n_1*\alpha^n_1]=[\beta_0^n*l^n*\alpha_0^n]=[\beta_0^n][l^n]
[\alpha_0^n]=[\beta_0^n][\alpha_0^n][l^n]=[\beta_0^n*\alpha_0^n][l^n]=
[\beta^n_1*\alpha^n_1][l^n].\] 
In the third step we have exploited the fact that a topological
group always has commutative fundamental group \cite{pos}. This shows that
$[l^n]=1$ that is the loop $l^n$ is contractible in $G$ for all 
$0\leq n\leq k$ in other words the homotopy $F$ is deformable into a
homotopy $F'$ yielding
$f_0$ and $f_1$ are homotopic in $\left[ (W, x^1,\dots ,x^k, i^0),
(G,e)\right]$ as well. $\Diamond$
\vspace{0.1in}

\noindent The above proposition enables us to give a more explicit
description of the set (\ref{ujvakuum}).
\begin{theorem}
Let $(M,g)$ be a space-time obeying the null energy
condition. Assume it contains a single asymptotically
flat region with globally hyperbolic domain of outer communication.
Suppose this region contains a Cauchy surface homeomorphic to the interior
of a compact three-manifold $V$. Let $G$ be a typical compact Lie
group i.e., let $G$ be $U(n)$ with $n\geq 2$, or $SO(n)$, {\em Spin}$(n)$
with $n\not=4$, or $SU(n)$, $Sp(n)$ for all $n$, or
$G_2$, $F_4$, $E_6$, $E_7$, $E_8$. Then we have
\[\left[ (V, \partial V, i^0), (G, e)\right]\cong\Z .\]
Moreover we have  
\[\left[ (V, \partial V, i^0), (U(1), e)\right]\cong 0,\]
and
\[\left[ ( V, \partial V, i^0), (SO(4),
e)\right]\cong\left[ (V, \partial V, i^0), (\mbox{{\em Spin}}(4),
e)\right]\cong \Z\oplus\Z\]
for the remaining cases.
\label{fotetel}
\end{theorem}
{\it Proof.} In light of the above considerations and Proposition
\ref{homotopiapropozicio}, we have 
\[[( V, \partial V, i^0), (G,e)]\cong [(W,
x^1,\dots ,x^k, i^0), (G,e)]\cong [(W, i^0), (G,e)].\]
Hence we can use the results of Isham who classified the set $[(W,
i^0), (G,e)]$ and it is summarized in \cite{isham1} in Table 1 on p. 207.
But in our case $W$ is a connected, closed, simply connected
three-manifold hence the above result follows. $\Diamond$
\vspace{0.1in}

\noindent {\it Remark.} We mention that assuming the validity of the three
dimensional Poincar\'e conjecture i.e., if $W\cong S^3$, then our
theorem can be derived without using Isham's result since in this case we
have simply $[( V, \partial V, i^0), (G,e)]\cong\pi_3(G)$.

We can see by this result that although the homotopy set
(\ref{ujvakuum}) of effectively deformable vacua is typically non-trivial,
it is remarkable more simple than in the original calculations of Isham et
al. based on topological considerations only. The homotopy sets listed in
Theorem \ref{fotetel} are exactly the same as for the flat Minkowskian
case. Being all these vacua gauge equivalent (since $N$ is simply 
connected) we have to introduce again linear combinations like 
(\ref{teta}) in this more general situation. Consequently we can see 
that approach (i) to the $\theta$-parameter, mentioned in the Introduction, 
still makes sense in the general case.
 
\section{Conclusion and outlook}
In this paper we have studied the concept of $\theta$-vacua
in general Yang--Mills theories. In light of our results, we can see that
for outer observers in asymptotically flat space-times
$\theta$-vacua do occur in a Yang--Mills theory and they can be 
constructed in a rigorous way by referring to geometric quantization as 
follows. 

For simplicity we restrict attention to a simple gauge group $G$. For the 
moduli space ${\cal V}$ of classical vacuum solutions of a Yang--Mills 
theory over an asymptotically flat region $(N,g\vert_N)$ we have the 
identification ${\cal V}\cong\left[ (V, \partial V, i^0), (G, e)\right]$ 
as we have seen. Furtheromore Theorem \ref{fotetel} says that as a set we 
have a homeomorphism ${\cal V}\cong\Z$. This space, regarded as a noncompact 
zero dimensional manifold is naturally identified with its cotangent 
bundle $T^*{\cal V}$. This setup resembles the situation of a {\it 
real polarization} in geometric quantization. Within this framework then 
the Hilbert space of the vacuum sector of the corresponding quantum 
Yang--Mills theory is identified with the space of $L^2$ functions on 
${\cal V}$:
\[{\cal H}_N=\{ f: {\cal V}\rightarrow\C\:\vert\:\Vert f\Vert_{L^2({\cal 
V})}<\infty\} .\]
This means that an element $f\in{\cal H}_N$ is described by 
complex numbers $a_n$ with $-\infty <n<\infty$ satisfying simply
\[\sum\limits_{n=-\infty}^\infty\vert a_n\vert^2<\infty .\]
Assigning to this function $f$ the convergent Fourier series
\[f(\theta ):=\sum\limits_{n=-\infty}^\infty a_n{\rm e}^{{\bf i}n\theta}\]
we identify naturally this vacuum Hilbert space with the space of 
square integrable functions on the circle
\[{\cal H}(S^1):=\{ f:S^1\rightarrow\C\:\vert\:\Vert 
f\Vert_{L^2(S^1)}<\infty\} .\]
The isomorphism ${\cal H}_N\cong{\cal H}(S^1)$ provides a very 
straightforward, natural introduction of a $\theta$ parameter into the 
vacuum sector of Yang--Mills theories as was done heuristically in 
(\ref{teta}). Moreover we can see that the whole quantum vacuum is just 
linear combination of $\theta$-vacua. The generalization to non-simple 
gauge groups is clear. Observe however that in this picture the role of 
causality is extremely important: without it the classical moduli ${\cal V}$ 
considered over the whole original space-time $(M,g)$ would be complicated 
topologically in the sense that in general even the connected components 
of ${\cal V}$ would be non-zero dimensional manifolds with non-trivial
topology introducing some degeneracy into the vacuum sector.

In summary we can say that despite the possible 
complicated topology of the underlying Cauchy surface of an 
asymptotically flat space-time, the vacuum structure is similar to the 
flat Minkowskian case, due to the
causal structure of these space-times which is complicated in the general
case, too. Hence the introduction of the various new CP-violating terms
studied in \cite{isham5} are unnecessary. Taking seriously the causal 
structure experienced by an observer also fits with the {\it Heisenberg 
dictum} that quantum field theory should be formulated in terms of 
observers.

The suppression of the topology of the underlying Cauchy surface is due to
the result of Galloway or Chru\'sciel--Wald which is a consequence of
the so-called Topological Censorship Theorem of Friedman--Schleich--Witt
\cite{fri-sch}. Consequently, the reduction of the problem of the general
CP-violation to the flat Minkowskian case is essentially due to this
result. However Topological Censorship remains valid in a more general
(i.e., not only an asymptotically flat) setting \cite{gal-sch-wit-woo};
therefore we may expect that our attack on the first approximation of the
CP-problem may continue to hold in these more general situations.

Finally, natural questions arise: Are there instanton solutions in the
corresponding Wick-rotated theories? Recent results on constructing
$SU(2)$ instanton solutions over various gravitational instantons may
point towards this possibility \cite{ete2}\cite{ete-hau}. What 
is the physical relevance of these solutions? Do they induce 
semi-classical tunnelings between vacuum states of different effective 
winding numbers? If the answer for these questions is yes, beyond (i) we 
have another, more physical, reason to introduce $\theta$-vacua by concept 
(ii), also mentioned in the Introduction.
\vspace{0.1in}

\noindent{\bf Acknowledgement.} The author expresses his thanks to G.W. 
Gibbons for calling attention for Isham's papers and the stimulating 
comments in Japan six years ago. The work was partially supported by OTKA 
grants No. T43242 and No. T046365 (Hungary).


\begin{thebibliography}{99}

\bibitem{bpst}  A.A. Belavin, A.M. Polyakov, A.S. Schwarz,
Yu.S. Tyupkin: {\it Pseudoparticle solutions of the Yang--Mills 
equations}, Phys. Lett. {\bf B59}, 85-87 (1975);

\bibitem{che} L-P. Cheng, L-F.Li: Gauge Theory of Elementary
Particle Physics, Clarendon Press (1984);

\bibitem{chr-wal} P.T. Chru\'sciel, R.M. Wald: {\it On the topology of 
stationary black holes}, Class. Quant. Grav. {\bf 11}, L147-L152 (1994) ;

\bibitem{isham5} S. Deser, M.J. Duff, C.J. Isham: {\it Gravitationally 
induced CP-effects}, Phys. Lett. {\bf B93}, 419-423 (1980);

\bibitem{ete} G. Etesi: {\it The structure of Yang--Mills vacua seen by a 
distant observer}, in: Consistent equation of classical gravitation 
to quantum limit and beyond, ed.: Sidharth, B.G., 
Altaisky, M.V., Kluwer Academic/Plenum Press Publishers, New York (2001);

\bibitem{ete2} G. Etesi: {\it Classification of 't Hooft instantons over
multi-centered gravitational instantons}, Nucl. Phys. {\bf B662}, 511-530
(2003);

\bibitem{ete-hau} G. Etesi, T. Hausel: {\it On Yang--Mills instantons over 
multi-centered gravitational instantons}, Commun. Math. Phys. {\bf 235}, 
275-288 (2003);

\bibitem{fri-sch} J.L. Friedman, K. Schleich, D.M. Witt: {\it Topological 
censorship}, Phys. Rev. Lett. {\bf 71}, 1486-1489 (1993);

\bibitem{gal} G.J. Galloway: {\it On the topology of the domain of outer 
communication}, Class. Quant. Grav. {\bf 12}, L99-L101 (1995);

\bibitem{gal-sch-wit-woo} G.J. Galloway, K. Schleich, D.M. Witt, E.
Woolgar: {\it Topological censorship and higher genus black holes}, Phys. 
Rev. {\bf D60}, 104039, 11 pp. (1999);

\bibitem{gan} D. Gannon: {\it Singularities in nonsimply connected 
space-times}, Journ. Math. Phys. {\bf 16}, 2364-2367 (1975);

\bibitem{haw} S.W. Hawking: {\it Virtual black holes}, Phys. Rev. {\bf 
D53}, 3099-3107 (1996);

\bibitem{tho2} G. 't Hooft: {\it Symmetry breaking through Bell--Jackiw 
anomalies}, Phys. Rev. Lett. {\bf 37}, 8-11 (1976);

\bibitem{tho} G. 't Hooft: {\it How instantons solve the U(1) 
problem}, Phys. Rep. {\bf 142}, 357-387 (1986);

\bibitem{isham1} C.J. Isham: {\it Vacuum tunneling in static space-times}, 
in: Old and new Questions in Physics, Cosmology, Philosophy and 
Theoretical Biology, Ed.: A. Van Der Merwe, Plenum Press, New York, 
189-211 (1983);

\bibitem{isham3} C.J. Isham, G. Kunstatter: {\it Spatial topology and 
Yang--Mills vacua}, Journ. Math. Phys. {\bf 23}, 1668-1677 (1982);

\bibitem{isham4} C.J. Isham, G. Kunstatter: {\it Yang--Mills canonical 
vacuum structure in a three-space}, Phys. Lett. {\bf B102}, 417-420 
(1981);

\bibitem{kak} M. Kaku: Quantum Field Theory, Oxford University
Press, Oxford (1993);

\bibitem{pec} R.D. Peccei, H.R. Quinn: {\it Constraints imposed by CP 
conservation in the presence of pseudoparticles}, Phys. Rev. {\bf D16}, 
1791-1797 (1977);

\bibitem{pres-triv} J. Preskill, S.P. Trivedi, M.B. Wise: {\it Wormholes 
in spacetime and $\theta_{QCD}$}, Phys. Lett.
{\bf B223}, 26-31 (1989);

\bibitem{pos} M. Postnikov: Lectures on Geometry V.: Lie groups and Lie
algebras, Mir Publishers, Moscow (1987);

\bibitem{spa} E.H. Spanier: Algebraic Topology, Springer--Verlag, Berlin
(1966);

\bibitem{wal} R.M. Wald: General Relativity, Univ. of Chicago Press,
Chicago (1984);

\bibitem{wit} D.M. Witt: {\it Vacuum space-times that admit no maximal 
slice}, Phys. Rev. Lett. {\bf 57}, 1386-1389 (1986).

\end{thebibliography}
\end{document}